\documentclass[12pt,english]{article}
\usepackage[T1]{fontenc}
\usepackage[latin9]{inputenc}
\usepackage[letterpaper]{geometry}
\usepackage{amsmath}
\usepackage{setspace}
\usepackage{amssymb}
\makeatletter

\providecommand{\tabularnewline}{\\}

\usepackage{amsfonts}\usepackage{amsthm}\usepackage{commath}\usepackage{booktabs}\usepackage{lscape}\usepackage{xr}
\newcommand{\myref}[2]{\hyperref[#1]{#2}}\numberwithin{equation}{section}

\usepackage{latexsym}
\usepackage{enumerate}
\usepackage[toc,title,titletoc,header]{appendix}\usepackage[bottom]{footmisc}
\usepackage[pdfborder={0 0 0}]{hyperref}

\interfootnotelinepenalty=10000\usepackage{lineno}\setpagewiselinenumbers
\newtheorem{theorem}{Theorem}[section]\theoremstyle{definition}

\theoremstyle{remark}\newcounter{assumptionM}\newcounter{assumptionA}\def\theassumptionM{M.\arabic{assumptionM}}\def\theassumptionA{A.\arabic{assumptionA}}

\makeatother

\usepackage{babel}

\makeatother

\usepackage{babel}

\makeatother

\usepackage{babel}

\makeatother

\usepackage{babel}

\begin{document}
\sloppy

\title{An Upper Bound for Functions of Estimators in High Dimensions }

\author{\textsc{Mehmet Caner}%
\thanks{\textsc{North Carolina State University, Nelson Hall, Department of
Economics, NC 27695. Email: mcaner@ncsu.edu.}%
} \and \textsc{Xu Han}%
\thanks{\textsc{City University of Hong Kong, 83 Tat Chee Avenue, KL, Hong
Kong S.A.R. Email: xuhan25@cityu.edu.hk}%
}}

\date{\today}
\maketitle
\begin{abstract}
We provide an upper bound as a random variable for  the functions of estimators
in high dimensions. This upper bound may help establish the rate
of convergence of functions in high dimensions. The upper bound random
variable may converge faster, slower, or at the same rate as estimators
depending on the behavior of the partial derivative of the function.
We illustrate this via three examples. The first two examples use
the upper bound for testing in high dimensions, and third example
derives the estimated out-of-sample variance of large portfolios.
All our results allow for a larger number of parameters, $p$, than
the sample size, $n$. 

\noindent \textit{Keywords and phrases}: Many assets, many restrictions,
Lasso.


\end{abstract}


\newpage{}

\section{Introduction}

The delta method is one of the most widely used theorems in econometrics
and statistics. It is a very simple and a useful idea. It can provide
limits for complicated functions of estimators as long as the function
is differentiable. The limit of the function of estimators can be obtained from the limit of the estimators, with
 the same rate of convergence. In the case of finite-dimensional
parameter estimation, since the derivative at the parameter value
is finite, rates of convergence of both estimators and function of
estimators are the same.

In the case of high dimensional parameter estimation, we show that
this is not the case, and the rates of convergence may change. We
show that the structure of the derivative of the function is the key.
An upper bound random variable is provided for the functions of estimators
in high dimensions. We show this upper bound on functions of estimators
may converge faster, slower, or at the same rate as estimators. Even
though a new delta theorem is not provided, 
this upper bound can  get the rate of convergence of
functions of estimators. For example, the variance of a portfolio
is a quadratic function of the portfolio weight. Our theorem implies
that the convergence rate of the portfolio variance is slower than
that of the estimated weight when the number of assets is diverging
(see Example 3 in Section 3). From now on we denote the number of assets as $p_n$ since they grow with sample size.

Our result is useful when the number of parameters $p_{n}$ is larger
than the sample size $n$, where $p_{n}$ grows with $n$. The reason
is that proofs in high dimensional problems usually depend on knowing
the rate of convergence of the bound, which may be a lasso-type estimation,
oracle inequality, or problems in high dimensional portfolio analysis.

After the main theorem, we illustrate our point in three examples:
first by examining a linear function of estimators that is heavily
used in econometrics, second a new debiased lasso type of estimator,
and third by analyzing the out-of-sample variance of a large portfolio
of assets in finance.

Section 2 provides our theorem. Section 3 has three examples. Section
4 provides a discussion of how the results may be tied to nonparametric analysis
and many weak instrument asymptotics. Appendix shows the proof
and provides more examples.

\section{Upper Bound}

Let $\beta_{0}=(\beta_{1,0},\cdots,\beta_{p_n,0})'$ be a $p_n\times1$
parameter vector with an estimator $\hat{\beta}=(\hat{\beta}_{1},\cdots,\hat{\beta}_{p_n})'$.
Define a function $f(.)$, $f:K\subset R^{p_n}\to R^{m}$.  To be specific 
\textbf{$m$} is a constant unless noted otherwise.

We provide two conditions  for our theorem. First, denote
a column vector of $p_{n}$ zeros by $0_{p_{n}}$. Let $f_{d}(.)$
represent the $m\times p_{n}$ matrix of derivatives.

\textit{Condition C1. For all $h\neq0_{p_{n}}$, and $h$ is a $p_{n}\times1$
vector \[
\lim_{h\to0}\frac{\|f(\beta_{0}+h)-f(\beta_{0})-f_{d}(\beta_{0})h\|_{2}}{\|h\|_{2}}=0,\]
 where $\|.\|_{2}$ is the Euclidean norm for a generic vector, and
$f_{d}(\beta_{0})$ is an $m\times p_{n}$ matrix, whose $(i,j)$ th cell
consists of $\partial f_{i}/\partial\beta_{j}$ evaluated at $\beta_{0}$,
for $i=1,\cdots,m$, $j=1,\cdots,p_{n}$. }

Second, define the rate of convergence of an estimator as $r_{n}$,
a positive sequence in $n$, and when $n\to\infty$, $r_{n}\to\infty$.

\textit{Condition C2. \[
r_{n}\|\hat{\beta}-\beta_{0}\|_{2}=O_{p}(1).\]
 }

Condition C1 is a high-level assumption that shapes our function of
interest. It restricts the function $f$ to be differentiable. Hence,
C1 rules out continuous functions that are non-differentiable at $\beta_{0}$. Condition C2 gives a convergence
rate for the estimator of interest. Several examples about $\hat{\beta}$
and $r_{n}$ that satisfy C2 are provided in Section 3. 

The following theorem is the main theoretical result. Given the convergence
rate of the high-dimensional estimator, it  provides an upper
bound (random variable) for the estimation error of functions of the
high-dimensional estimator. This is useful in econometric theory since
it can give us an idea about what the rate of convergence of functions
of estimators might be. Let $k_{n}$ and $d_{n}$ be positive sequences
in \textbf{$n$} so that both $k_{n}\to\infty$ and $d_{n}\to\infty$
as $n\to\infty$. Take a generic matrix, $D$,  of dimensions $ m \times p_n$:  $\|| D \||_{2}$
denotes the Frobenius norm of the matrix $D$: 
\[ \|| D \||_2 = \sqrt{\sum_{i=1}^m \| d_i \|_2^2},\]
where $d_i$ is a $p_n \times 1$ vector, and its transpose $d_i'$ is the $i$th row of $D$.

\begin{theorem} Let Conditions C1 and C2 hold, and $C>0$ be a universal
constant. Assume that $\||f_{d}(\beta_{0})\||_{2}>0$

a) If $\||f_{d}(\beta_{0})\||_{2}=C$, then \begin{equation}
\|f(\hat{\beta})-f(\beta_{0})\|_{2}\le L_{n}=O_{p}(\frac{1}{r_{n}}).\label{3}\end{equation}

b) If $\||f_{d}(\beta_{0})\||_{2}=Ck_{n}$, then \begin{equation}
\|f(\hat{\beta})-f(\beta_{0})\|_{2}\le L_{n}=O_{p}(\frac{k_{n}}{r_{n}}).\label{4b}\end{equation}

c) If $\||f_{d}(\beta_{0})\||_{2}=\frac{C}{d_{n}}$, then

\begin{equation}
\|f(\hat{\beta})-f(\beta_{0})\|_{2}\le L_{n}=O_{p}(\frac{1}{d_{n}r_{n}})+o_{p}(\frac{1}{r_{n}})=o_{p}(\frac{1}{r_{n}}).\label{4}\end{equation}

\label{thm1} \end{theorem}

\textbf{Remarks}. 1. The theorem does not provide a limit for functions
of estimators, so this is not the delta theorem.

2. Part (a) shows under what conditions we can get the same rate of
convergence for the functions of estimators compared with the rate
of convergence of $\hat{\beta}-\beta_{0}$. Example 2 illustrates
this point.

3. Part (b) shows that the $l_{2}$ norm of the partial derivative
function may change with the dimension of the parameter vector. When $r_n > k_n$, and 
$k_{n}\to\infty$, then the upper bound $L_{n}$ converges at a slower
rate than estimators of parameters. With $r_{n}/k_{n}\to0$, then
the upper bound is diverging which tells us that the function of estimators
may diverge, too.

4. Part (c) shows that the function of the estimators converges to
zero in probability faster than the rate of convergence
of estimators $r_{n}$.

5. Theorem \ref{thm1} also holds with $l_{1}$ and $l_{\infty}$ norms.
These new norm results can be shown when Conditions C1 and C2 hold in $l_{1}$ and $l_{\infty}$
norms. We discuss this in Part A of the Appendix. 


\section{Examples}

We now provide three examples that highlight the contribution. The
first one is related to the linear functions of estimators, the second
one considers the Debiased Conservative Lasso (DCL) of Caner and Kock (2018),
and the third one is related to the out-of-sample variance of large portfolios.

\textbf{Example 1}.

This example considers lasso, which is one of the benchmark methods
in machine learning. It is a penalized least squares estimator with
$l_{1}$ penalty. The penalty induces sparsity in the model, which
can prevent overfitting (Hastie, Tibshirani, Friedman, 2009, Tibshirani, 1996).

Let us denote $\beta_{0}$ as the true value of vector ($p_{n}\times1$)
of coefficients. The number of the true nonzero coefficients is denoted
by $s_{0}$, and $s_{0}>0$. A simple linear model is: \begin{equation}
y_{t}=x_{t}'\beta_{0}+u_{t},\label{eq:linear model}\end{equation}
for $t=1,\cdots,n$. For simplicity, we assume that $u_{t}$ are iid
errors with zero mean and finite variance and $x_{t}$ is a set of
$p_{n}$ deterministic regressors. The iid assumption on $u_{t}$
is to keep our illustration simple and the following result still
holds under more general assumptions on $u_{t}$ and $x_{t}$ as shown in Assumption 1 of 
Caner and Kock (2018). 

The lasso estimator is defined as \begin{equation}
\hat{\beta}=argmin_{\beta\in R^{p_{n}}}\left[\sum_{t=1}^{n}\frac{(y_{t}-x_{t}'\beta)^{2}}{n}+2\lambda_{n}\sum_{j=1}^{p_{n}}|\beta_{j}|\right],\label{eq:lasso}\end{equation}
 where $\beta_{j}$ is the $j$ th element of $\beta$, $\lambda_{n}$
is a positive tuning parameter, and it is established that $\lambda_{n}=O(\sqrt{\frac{logp_{n}}{n}})$.
Corollary 6.14 or Lemma 6.10 of Buhlmann and van de Geer (2011) shows
that, for lasso estimators $\hat{\beta}$, with $p_{n}>n$ \begin{equation}
r_{n}\|\hat{\beta}-\beta_{0}\|_{2}=O_{p}(1),\label{ex11}\end{equation}
 where \begin{equation}
r_{n}=\sqrt{\frac{n}{logp_{n}}}\frac{1}{\sqrt{s_{0}}}.\label{ex11a}\end{equation}

Given (\ref{ex11}), we may be interested 
in the large sample behavior of $D(\hat{\beta}-\beta_{0})$, where $D$ is an
$m\times p_{n}$ matrix. The $D$ matrix can be thought of putting
restrictions on $\beta_{0}$. We want to see whether $D(\hat{\beta}-\beta_{0})$
has a different rate of convergence from $\hat{\beta}-\beta_{0}$.
From our Theorem \ref{thm1}(a), it is clear that $f_{d}(\beta_{0})=D$.
Basically in the case of inference, this matrix and the vectors show how
many of $\beta_{0}$ will be involved with the restrictions. If we want
to use $s_{0}$ elements in each row of $D$ to test $m$ restrictions,
then $\||D\||_{2}=O(\sqrt{s_{0}})$. Note that this corresponds to
using $s_{0}$ elements in $\beta_{0}$ for testing $m$ restrictions.
In other words, if $k_{n}=s_{0}$ and $s_{0}\to\infty$ as $n\to\infty$,
then Theorem \ref{thm1}(b) implies that \[
L_{n}=O_{p}(\frac{\sqrt{s_{0}}}{r_{n}}).\]
 This means that even with a fixed number of $m$ restrictions, the
upper bound random variable has a slower rate of convergence than the
estimators, so it is possible that functions of estimators also converge
slower to a limit. 

\textbf{Remark}. In high dimensions, a common assumption is to impose
$\|d_{i}\|_{2}=1$. See, for example, Caner, Han and Lee (2018) and
Caner and Kock (2018). In that case, \[
\||D\||_{2}=\sqrt{m},\]
 where $m$ is fixed, but $p_{n}$ is growing with $n$. The rate of convergence of $D(\hat{\beta}-\beta_{0})$
will be still $r_{n}$, which is in (\ref{ex11a}). Thus, there will
be no slowdown of the rate of convergence and this is a sharp rate
since $\||f_{d}(\beta_{0})\||_{2}=\||D\||_{2}=\sqrt{m}$. This is
not an upper bound on $f_{d}(\beta_{0})$.

\textbf{Example 2.} Another estimator that is recently analyzed in
the context of $p_{n}>n$ is the DCL of Caner
and Kock (2018). Consider the model in example 1 above in the matrix form:
 \[
Y=X\beta_{0}+u,\]
 where $Y$ is an $n\times1$ vector, $X$ is an $n\times p_{n}$
matrix, $\beta_{0}$ is a $p_{n}\times1$ vector that consists of
$s_{0}$ nonzero parameters, $u$ is an $n\times1$ error vector.
$\hat{\beta}_{CL}$ is the conservative lasso estimator defined as
\[
\hat{\beta}_{CL}=argmin_{\beta\in R^{p_{n}}}\{\|Y-X\beta\|_{2}^{2}/n+2\lambda_{n}\sum_{j=1}^{p_{n}}\hat{w}_{j}|\beta_{j}|\},\]
 where $\hat{w}_{j}=\lambda_{n}/max(|\hat{\beta}_{j}|,\lambda_{n})$,
$max(a,b)$ chooses maximum of two elements $a$  or $b$, and $\hat{\beta}_{j}$
is the lasso estimator defined in example 1. The DCL is uniformly consistent, and it has a standard normal asymptotic
limit and an asymptotically valid uniform confidence band, unlike the lasso
and the conservative lasso. These are established in Theorem 3 of Caner
and Kock (2018). The formula for the DCL estimator $\hat{b}$ is: \[
\hat{b}=\hat{\beta}_{CL}+\hat{\Theta}X'(Y-X\hat{\beta}_{CL})/n,\]
 where $\hat{\Theta}$ is an approximate estimate for precision matrix,
which will be abstracted away in this paper. Information and detailed
properties are described in Section 3.2 of Caner and Kock (2018).
Specifically, Caner and Kock (2018) derive the rate of convergence
for Wald and $\chi^{2}$ type of tests. They also show that the confidence
bands on DCL are contracting at the optimal rate of $n^{-1/2}$. By
Theorem 2 of Caner and Kock (2018), we have \[
n^{1/2}(\hat{b}_{j}-\beta_{0j})=O_{p}(1),\]
 for $j=1,\cdots,p_{n}$. If we want to test $m$ restrictions on
$\beta_{0}$, with $H=[h_{1},...,h_{m}]^{\prime}$ representing the
restriction matrix of $m\times p_{n}$ dimension, then \[
\||H\||_{2}=\sqrt{\sum_{i=1}^{m}\|h_{i}\|_{2}^{2}}=\sqrt{m},\]
 given $\|h_{i}\|_{2}=1$ in Theorem 2 of Caner and Kock (2018). Conditions
C1 and C2 are satisfied. If $m$ is a fixed number, then $H(\hat{\beta}-\beta_{0})$
 converges to the limit  at rate $n^{1/2}$, and a $\chi^{2}$ type test 
converges to the limit  at rate $n$ as in (21) of Caner and Kock (2018).

\textbf{Example 3}. One of the main issues in finance is the analysis
of portfolio variance. If we denote the portfolio allocation vector
by $w$ ($p_{n}\times1$), and the covariance matrix of asset returns
by $\Sigma$, then the portfolio variance is $w'\Sigma w$. The out-of-sample
estimate of this portfolio variance is $\hat{w}'\Sigma\hat{w}$. This estimate
 can be seen in Ledoit and Wolf (2017) and Ao et al. (2019). The number
of assets, $p_{n}$, grows with $n$, which is the time span of the
portfolio, and $p_{n}>n$. Let $Eigmax(A)$ denote the maximum eigenvalue
of a matrix $A$ and $C>0$ be a positive constant.

We analyze the global minimum portfolio weights. Define $1_{p_{n}}$ as
the $p_{n}$ vector of ones. The weights are computed as

\[
w=\frac{\Sigma^{-1}1_{p_{n}}/p_{n}}{1_{p_{n}}'\Sigma^{-1}1_{p_{n}}/p_{n}}.\]

\noindent From Theorem 3.3 of Callot et al. (2019), the estimated
weights are:

\[
\hat{w}=\frac{\hat{\Theta}1_{p_{n}}/p_{n}}{1_{p_{n}}'\hat{\Theta}1_{p_{n}}/p_{n}},\]
 where $\hat{\Theta}$ is the nodewise regression estimate of $\Sigma^{-1}$.
Take $\beta_{0}=w$, and $\hat{\beta}=\hat{w}$. So our parameter
is of dimension $p_{n}$, and it is growing with $n$ and larger than
$n$. Our interest centers on the out-of sample portfolio variance
estimation, given that we can estimate weights consistently, with
a known rate of convergence. First, we start with verifying Condition
C1. See that $f(\hat{\beta})=\hat{\beta}'\Sigma\hat{\beta}$, and
$f(\beta_{0})=\beta_{0}'\Sigma\beta_{0}$.
Condition C2 also holds for this example, and we will show
that after verifying Condition C1.

For C1, we have

\[
\|f(\beta_{0}+h)-f(\beta_{0})-f_{d}(\beta_{0})h\|_{1}=\|(\beta_{0}+h)'\Sigma(\beta_{0}+h)-\beta_{0}'\Sigma\beta_{0}-2\beta_{0}'\Sigma h\|_{1}=\|h'\Sigma h\|_{1}.\]
 Next \begin{eqnarray*}
\frac{\|h'\Sigma h\|_{1}}{\|h\|_{1}} & = & \frac{|h'\Sigma h|}{\|h\|_{1}}\le\frac{\|h\|_{2}^{2}Eigmax(\Sigma)}{\|h\|_{1}}\\
 & \le & \frac{\|h\|_{2}^{2}}{\|h\|_{2}}Eigmax(\Sigma)=\|h\|_{2}Eigmax(\Sigma)\\
 & \to & 0\ \mathrm{as}\ \|h\|_{1}\to0,\end{eqnarray*}
 where the second inequality follows from $\|h\|_{1}\ge\|h\|_{2}$,
and for the convergence to zero we use  assumption $Eigmax(\Sigma)\le C<\infty$,
and the fact that $\|h\|_{1}\to0$ implies $\|h\|_{2}\to0$. Thus,
Condition C1 is satisfied. Then Condition C2 is satisfied since, by
Theorem 3.3 of Callot et al. (2019) we have \[
r_{n}\|\hat{w}-w\|_{1}=O_{p}(1),\]
 where $r_{n}=\frac{\sqrt{n}}{\sqrt{logp_{n}}}\frac{1}{\bar{s}^{3/2}}$,
with $\bar{s}=\max_{1\le j\le p}s_{j}$, and $s_{j}$ is the number
of nonzero cells in the $j$ th row of the precision matrix. Now, we derive
the rate of convergence of the out-of-sample variance estimator. Note
that \[
\|f_{d}(\beta_{0})\|_{1}=\|2\beta_{0}'\Sigma\|_{1}=2\|\Sigma w\|_{1},\]
 since $\Sigma$ is a $p_{n}\times p_{n}$ symmetric matrix and $\beta_{0}=w$.
Define $\sigma_{i,j}$ as the $(i,j)$ th element of  $\Sigma$
matrix. By (\ref{r2}) \begin{eqnarray*}
\|\Sigma w\|_{1} & \le & \||\Sigma\||_{1}\|w\|_{1}\\
 & = & [\max_{1\le j\le{p_{n}}}\sum_{i=1}^{p_{n}}|\sigma_{i,j}|]\|w\|_{1}\\
 & = & [\max_{1\le j\le{p_{n}}}\sum_{i=1}^{p_{n}}|\sigma_{i,j}|]O(\sqrt{\bar{s}})\\
 & = & O(\sqrt{\bar{s}}),\end{eqnarray*}
 where the third line uses $\|w\|_{1}=O(\sqrt{\bar{s}})$ by Theorem
3.3 of Callot et al. (2019), and the last line uses the assumption
$[\max_{1\le j\le{p_{n}}}\sum_{k=1}^{p_{n}}|\sigma_{i,j}|]\le C<\infty$.
Clearly, we can define $k_{n}:=O(2\|\Sigma w\|_{1})$, which implies
$k_{n}=O(\sqrt{\bar{s}})$ by the inequality above. Applying Theorem
\ref{thm1}(b), we obtain a slow rate for the out-of-sample variance
estimator compared with the weight estimation\[
\frac{\sqrt{n}}{\sqrt{logp_{n}}}\frac{1}{\bar{s}^{2}}|\hat{w}'\Sigma\hat{w}-w'\Sigma w|=O_{p}(1),\]
when  $\bar{s}$ is growing
with $n$.

\section{Discussion of the Upper Bound}

Here we provide a brief discussion about our results in  some
econometrics problems. We can see three areas related to our technique
that may be beneficial to the researchers. The first area is the many
weak instruments literature. In the case of many weak instruments
as in Newey and Windmeijer (2009) and Caner (2014), we can derive
the rate of convergence for the estimation of the sample moments from
the estimation of the parameters in the structural equation. If we have
$\hat{\beta}$ as the generalized empirical likelihood estimator,
with $r_{n}\|\hat{\beta}-\beta_{0}\|_{2}=O_{p}(1)$, then we can have
$r_{n}d_{n}\|\hat{g}(\hat{\beta})-\hat{g}(\beta_{0})\|_{2}=O_{p}(1)$,
with \[
\hat{g}(\hat{\beta})=\frac{1}{n}\sum_{i=1}^{n}Z_{i}(y_{i}-x_{i}'\hat{\beta}),\]
 where $Z_{i}$ is an $m\times1$ vector of instruments, $m$ is growing
with $n$. Let $y_{i}$ be  the outcome variable, and $x_{i}$ represent the control
and endogenous variables ($p_{n}\times1$ vector). This satisfies
our Theorem \ref{thm1}(c). The details of this example are provided
in the appendix. This extends the results of Example 1 in Section
2 of Caner (2014), and the linear model of Newey and Windmeijer (2009,
p. 690-698).


The second one is the portfolio analysis, where we illustrate our results
through an example in the main text. Since there are a lot of nonlinear
functions in parameters of interest, and it is neither obvious nor trivial
to get the rates of these functions in the modern portfolio theory when
the number of assets, $p_{n}$, is larger than the time span of the
portfolio. Our technique can help. It analyzes the partial derivative
of the function at the parameter and automatically finds the rate
of convergence through that.

The third area is nonparametric estimation. Our theorem can be applied
to obtain the convergence rate of the estimate of the  nonparametric function.
Consider, for example, the series estimation of the following model
\[
y_{i}=g(x_{i})+\varepsilon_{i},\ \mathrm{with}\ E(\varepsilon_{i}|x_{i})=0,\]
 where the unknown function $g(x)$ can be approximated by a linear
combination of basis functions $h^{p_{n}}(x)$ and $p_{n}$ is increasing
with $n$. Let $\mathcal{S}$ be the compact support of $x$. Assume
that \begin{equation}
\sup_{x\in\mathcal{S}}|g(x)-h^{p_{n}}(x)^{\prime}\beta|=O(p_{n}^{-\alpha})\ \mathrm{for\ some}\ \alpha>0.\label{eq:k^-a}\end{equation}
 Define $\beta^{p_{n}}\equiv\arg\min_{\beta}\sup_{x\in\mathcal{S}}|g(x)-h^{p_{n}}(x)^{\prime}\beta|.$
The estimator of $\beta^{p_{n}}$, denoted by $\hat{\beta}$, is obtained
by a linear regression of $y_{i}$ on $h^{p_{n}}(x_{i})$. Newey (1997)
shows that $\|\hat{\beta}-\beta^{p_{n}}\|_{2}=O_{p}\left(\sqrt{p_{n}}/\sqrt{n}+p_{n}^{-\alpha}\right)$,
which gives the rate for Condition C2. For Condition C1, consider
the linear function $f(\beta)=h^{p_{n}}(x_{0})^{\prime}\beta$ for
a given $x_{0}\in\mathcal{S}$, so $f(\beta^{p_{n}})$ is the approximation
of $g(x_{0})$ and $f(\hat{\beta})$ is the series estimator $h^{p_{n}}(x_{0})^{\prime}\hat{\beta}$.
We are interested in the convergence rate of $|f(\hat{\beta})-f(\beta^{p_{n}})|$.
Since $f_{d}(\beta^{p_{n}})=h^{p_{n}}(x_{0})^{\prime}$, Condition
C1 holds automatically. By our Theorem 2.1(b) we have \begin{equation}
|f(\hat{\beta})-f(\beta^{p_{n}})|\le\zeta(p_{n})\cdot O_{p}\left(\sqrt{\frac{p_{n}}{n}}+\frac{1}{p_{n}^{\alpha}}\right)=O_{p}\left(\zeta(p_{n})(\sqrt{p_{n}}/\sqrt{n}+p_{n}^{-\alpha})\right),\label{eq:rate}\end{equation}
 where $\zeta(p_{n})=\sup_{x\in\mathcal{S}}\|h^{p_{n}}(x)\|_{2}$.%
\footnote{Under some regularity conditions, it follows that $\zeta(p_{n})=O(p_{n})$
for power series and $\zeta(p_{n})=O(\sqrt{p_{n}})$ for spline series
(Corollary 15.1, Li and Racine, 2007). %
} Together with (\ref{eq:k^-a}), the rate in (\ref{eq:rate}) implies
\[
|h^{p_{n}}(x_{0})^{\prime}\hat{\beta}-g(x_{0})|\le|f(\hat{\beta})-f(\beta^{p_{n}})|+\sup_{x\in\mathcal{S}}|h^{p_{n}}(x)^{\prime}\beta^{p_{n}}-g(x)|=O_{p}\left(\zeta(p_{n})(\sqrt{p_{n}}/\sqrt{n}+p_{n}^{-\alpha})\right),\]
 which is the well known convergence rate for the series estimator.
Note that our theorem actually gives a sharp upper bound on the convergence
rate of the series estimator.

\section{Simulation}

In this section, we study the degree of conservativeness  of  the bound derived by
Theorem \ref{thm1} via simulation. We consider the lasso estimator
discussed in Example 1. The model is generated using \eqref{eq:linear model},
where $x_{i}\sim iid\ N(0,I_{p_{n}})$, $u_{i}\sim iid\ N(0,s_{0})$,
and $\beta_{0}=(\mathbf{1}_{1\times s_{0}},0_{1\times(p_{n}-s_{0})})^{\prime}$.
We set $s_{0}\in\{5,10\}$, $p_{n}\in\{50,100,200,300\}$, and $n\in\{100,200,300\}$.
Since $\lambda_{n}=O(\sqrt{\log(p_{n})/n})$, we select the optimal
tuning parameter from the set $\{\lambda_{n}=c\sqrt{\log(p_{n})/n},\ c=0.1,0.25,0.5,1,2,3,4,5,6,7,8,9,10\}$
by minimizing the following information criterion, \[
\lambda^{*}=\arg\min_{\lambda_{n}}[\log\hat{\sigma}^{2}(\lambda_{n})+\frac{\hat{s}(\lambda_{n})}{n}\log(n)\log\log(p_{n})],\]
where $\hat{s}(\lambda_{n})$ is the number of nonzero entries in
the lasso estimator given by \eqref{eq:lasso} using tuning parameter
$\lambda_{n}$, and $\hat{\sigma}^{2}(\lambda_{n})$ is the corresponding
mean squared residuals. The term $\log\log(p_{n})$ follows the design
of Caner, Han and Lee (2018) to deal with the high dimensionality. 

We focus on the inference of the first $s_{0}$ parameters in $\beta_{0}$,
so $f(\beta_{0})=D\beta_{0}$ with $D=(I_{s_{0}},0)$. We compute
the ratio $\||f_{d}(\beta_{0})\||_{2}\|\hat{\beta}-\beta_{0}\|_{2}/\|f(\hat{\beta})-f(\beta_{0})\|_{2}=s_{0}\|\hat{\beta}-\beta_{0}\|_{2}/\|D\hat{\beta}-D\beta_{0}\|_{2}$
to see the tightness of the bound. Table 1 reports the average of
this ratio over 1000 replications. It is evident that the averaged
ratio is very close to $s_{0}$, which implies that most of the zero elements
in $\beta_{0}$ are estimated as zero by the lasso. In the (infeasible)
oracle case where all zero parameters are exactly estimated as zero,
the ratio would be equal to $s_{0}$. 
 We clearly see that our upper bound grows with sparsity, and will not be that tight unless the model is very sparse. However, the main problem is that in high dimensional econometrics, it is complicated to get 
 closed-form solutions; hence we rely on upper bounds. For example, the oracle inequalities, $l_1$ norm results  are all upper bounded.

\vspace{1em}

\noindent \begin{center}
Table 1: The Ratio of the Upper Bound to the Function of the Lasso Estimator \vspace{1em}

\par\end{center}

\noindent \begin{center}
\begin{tabular}{|c|c|c|c|c|}
\hline 
$n,\ s_{0}$ & $p_{n}=50$ & $p_{n}=100$ & $p_{n}=200$ & $p_{n}=300$\tabularnewline
\hline
\hline 
100, 5 & 5.23 & 5.09 & 5.04 & 5.02\tabularnewline
\hline 
200, 5 & 5.20 & 5.11 & 5.06 & 5.05\tabularnewline
\hline 
300, 5 & 5.15 & 5.08 & 5.06 & 5.04\tabularnewline
\hline 
100, 10 & 10.26 & 10.11 & 11.64 & 13.40\tabularnewline
\hline 
200, 10 & 10.45 & 10.20 & 10.09 & 10.07\tabularnewline
\hline 
300, 10 & 10.41 & 10.22 & 10.13 & 10.10\tabularnewline
\hline
\end{tabular}
\par\end{center}

\section{Conclusion}

We provide an upper bound for the functions of the estimators in high
dimensions. We also show three examples to illustrate the
main theorem. It is possible to extend our result to more relevant econometrics
issues in the age of big data. In summary, our method can be useful
for obtaining the rate of convergence of functionals of estimators,
since it uses the partial derivative of the function at $\beta_{0}$. Our bound can be beneficial in high dimensional
scenarios where it may be difficult to have direct proof of the rate
of convergence.\\


\textbf{REFERENCES}

Abadir, K. and J.R. Magnus (2005). Matrix Algebra. Cambridge University
Press. Cambridge.

Ao, M., Y.Li and X. Zheng (2019). Approaching mean-variance efficiency
for large portfolios. \textit{Review of Financial Studies}, Forthcoming.

Buhlmann, P. and S. van de Geer (2011). Statistics for High-Dimensional
Data. Springer Verlag, Berlin.

Callot, L., Caner, M., O, Onder, E. Ulasan (2019). A nodewise regression
approach to estimating large portfolios. \textit{Journal of Business
and Economic Statistics}, Forthcoming.

Caner, M. (2014). Near Exogeneity and Weak Identification in Generalized
Empirical Likelihood Estimators. \textit{Journal of Econometrics},
182, 247-288.

Caner, M, and X. Han, and Y. Lee (2018). Adaptive Elastic Net GMM
Estimation with Many Invalid Moment Conditions: Simultaneous Model
and Moment Selection. \textit{Journal of Business and Economics Statistics},36, 24-46.

Caner, M. and A.B. Kock. (2018). Asymptotically Honest Confidence
Regions for High Dimensional Parameters by the Desparsified Conservative
Lasso. \textit{Journal of Econometrics}, 203, 143-168.

Fan, J., Y. Liao, and X. Shi (2015). Risks of large
portfolios. \textit{Journal of Econometrics}, 186,
367-387.

Hastie,  T., R. Tibshirani, and J. Friedman (2009). The Elements of Statistical Learning. Springer Verlag. NYC.

Horn, R.A. and C. Johnson (2013). Matrix Analysis. Second Edition.
Cambridge University Press, Cambridge.

Ledoit, O., M. Wolf (2017). Nonlinear shrinkage of the covariance
matrix for portfolio selection: Markowitz meets Goldilocks. \textit{Review
of Financial Studies}, 30, 4349-4388.

Li, Q. and J. Racine (2007). Nonparametric Econometrics: Theory and
Practice. Princeton University Press.

Newey, W, (1997). Convergence Rates and Asymptotic Normality for Series
Estimators. \emph{Journal of Econometrics}, 79(1), 147-168.\textbf{ }

Newey, W, and F. Windmeijer (2009). Generalized Method of Moments
with Many Weak Moment Conditions. \textit{Econometrica}, 77, 687-719.

Tibshirani, R. (1996). Regression Shrinkage and Selection via the
Lasso. \emph{Journal of the Royal Statistical Society Series B}, 58,
267-288. 

Van de Geer, S., P. Buhlmann, Y. Ritov, and R. Dezeure (2014). On
asymptotically optimal confidence regions and test for high-dimensional
models. \textit{The Annals of Statistics,} 42, 1166-1202.

Van der Vaart, A.W. (2000). Asymptotic Statistics. Cambridge University
Press, Cambridge.

\section{Appendix}


The appendix has three parts. In Part A, we introduce the matrix inequalities
that are used  in the proofs. In Part B, we provide the proof
of our theorem and show why a classical proof of the delta method
in high dimensions does not work. In Part C, we give more examples
that are tied to our main theorem. \\

PART A.\\

Take a generic matrix, $A$, which is of dimension $m\times p_{n}$.
Denote the Frobenius norm for a matrix as $\||A\||_{2}=\sqrt{\sum_{i=1}^{m}\sum_{j=1}^{p_{n}}a_{ij}^{2}}$.
Note that in some literature such as Horn and Johnson (2013),
this definition is not considered a matrix norm, due to the lack of submultiplicativity.
However, our results will not change regardless of matrix norm definitions.
If we use Horn and Johnson (2013) definitions, our results can be
summarized in an algebraic form, rather than the matrix norm format. Define
\[
A=\left[\begin{array}{c}
a_{1}'\\
\vdots\\
a_{m}'\end{array}\right],\]
 where $a_{i}$ is a $p_{n}\times1$ vector, and its transpose is
$a_{i}'$, $i=1,\cdots,m$. Then for a generic $p_{n}\times1$ vector
$x$, \begin{equation}
\|Ax\|_{2}=\sqrt{\sum_{i=1}^{m}(a_{i}'x)^{2}}\le\left(\sqrt{\sum_{i=1}^{m}\|a_{i}\|_{2}^{2}}\right)\|x\|_{2}=\||A\||_{2}\|x\|_{2},\label{1}\end{equation}
 where the inequality is obtained by the Cauchy-Schwarz inequality. Note
that if we apply Horn and Johnson's (2013) norm definition, this matrix
norm inequality still holds, but we cannot use the matrix norm. In
that case we have \begin{equation}
\|Ax\|_{2}\le\left(\sqrt{\sum_{i=1}^{m}\|a_{i}\|_{2}^{2}}\right)\|x\|_{2}.\label{2}\end{equation}

Also, see that the results (\ref{3})-(\ref{4}) can be obtained in
other matrix norms.  A simple Holder's inequality provides \begin{equation}
\|Ax\|_{1}\le\||A\||_{1}\|x\|_{1},\label{r2}\end{equation}
 where we define the maximum column sum matrix norm: $\||A\||_{1}=\max_{1\le j\le p_{n}}\sum_{i=1}^{m}|a_{ij}|$,
and $a_{ij}$ is the $(i,j)$ th element of the matrix $A$. Theorem \textbf{\ref{thm1}} can be written in $l_{1}$ norm replacing
$l_{2}$ norm. We can also extend these results to another norm. A
simple inequality provides \begin{equation}
\|Ax\|_{\infty}\le\||A\||_{\infty}\|x\|_{\infty},\label{r4}\end{equation}
 where we define the maximum row sum matrix norm: $\||A\||_{\infty}=\max_{1\le i\le m}\sum_{j=1}^{p_{n}}|a_{ij}|$
and  Theorem \ref{thm1} can be written in $l_{\infty}$ norm as well.

\vspace{1in}

PART B.\\

First, we show why the classical proof of the delta method does not work
in high dimensions. However, this is not a negative result since it
 guides us towards the solution.

First by Condition C1, via p.352 of Abadir and Magnus (2005) $l(.):D\subset R^{p_{n}}\to R^{m}$
is a vector function \begin{equation}
\|f(\hat{\beta})-f(\beta_{0})-f_{d}(\beta_{0})[\hat{\beta}-\beta_{0}]\|_{2}=\|l(\hat{\beta}-\beta_{0})\|_{2}.\label{pt1}\end{equation}
 and \begin{equation}
\|l(\hat{\beta}-\beta_{0})\|_{2}=o_{p}(\|\hat{\beta}-\beta_{0}\|_{2}),\label{pt2}\end{equation}
 where we use Lemma 2.12 of van der Vaart (2000).

Since we are given $r_{n}\|\hat{\beta}-\beta_{0}\|_{2}=O_{p}(1)$
which is Condition C2, by (\ref{pt2}) \begin{equation}
r_{n}\|l(\hat{\beta}-\beta_{0})\|_{2}=o_{p}(1).\label{pt4}\end{equation}
 By (\ref{pt1})-(\ref{pt4}) \begin{equation}
\|r_{n}[f(\hat{\beta})-f(\beta_{0})]-r_{n}[f_{d}(\beta_{0})][\hat{\beta}-\beta_{0}]\|_{2}=o_{p}(1).\label{pt5}\end{equation}
 But this is the same result as in the regular delta method. (\ref{pt5})
is mainly a simple extension of Theorem 3.1 in van der Vaart (2000)
to Euclidean spaces so far. However, the main caveat comes from the derivative
matrix $f_{d}(\beta_{0})$, which is of dimension $m\times p_{n}$.
The rate of the matrix plays a role when $p_{n}\to\infty$ as $n\to\infty$.
For example, both $r_{n}[f_{d}(\beta_{0})][\hat{\beta}-\beta_{0}]$
and $r_{n}[f(\hat{\beta})-f(\beta_{0})]$ may be diverging, but $r_{n}\|\hat{\beta}-\beta_{0}\|_{2}=O_{p}(1)$.
Hence the delta method is not that useful if our interest centers
on getting rates for estimators as well as functions of estimators
that converge. In the fixed $p$ case, this is not an issue, since
the matrix derivative will not affect the rate of convergence at all,
as long as this is bounded away from zero, and is bounded from above away from infinity.
Note that boundedness assumptions may not be intact when we have $p_{n}\to\infty$,
as $n\to\infty$. Next part shows how to correct this problem.

\textbf{Proof of Theorem \ref{thm1}}. From Condition C1, using p.352
of Abadir and Magnus (2005), or proof of Theorem 3.1 in van der Vaart
(2000) yields

\[
f(\hat{\beta})-f(\beta_{0})=f_{d}(\beta_{0})[\hat{\beta}-\beta_{0}]+l(\hat{\beta}-\beta_{0}).\]

\noindent  Using the  Euclidean norm and the triangle inequality,
we have \begin{eqnarray*}
\|f(\hat{\beta})-f(\beta_{0})\|_{2} & = & \|f_{d}(\beta_{0})[\hat{\beta}-\beta_{0}]+l(\hat{\beta}-\beta_{0})\|_{2}\\
 & \le & \|f_{d}(\beta_{0})[\hat{\beta}-\beta_{0}]\|_{2}+\|l(\hat{\beta}-\beta_{0})\|_{2}.\end{eqnarray*}
 Next, multiply each side by $r_{n}$, and use (\ref{pt4}) \begin{equation}
r_{n}\|f(\hat{\beta})-f(\beta_{0})\|_{2}\le r_{n}\|f_{d}(\beta_{0})[\hat{\beta}-\beta_{0}]\|_{2}+o_{p}(1).\label{pt6}\end{equation}
 Then apply the matrix norm inequality in (\ref{1}) to the first term
on the right side of (\ref{pt6}) \begin{equation}
r_{n}\|f_{d}(\beta_{0})[\hat{\beta}-\beta_{0}]\|_{2}\le r_{n}\left[\||f_{d}(\beta_{0})\||_{2}\right]\left[\|\hat{\beta}-\beta_{0}\|_{2}\right].\label{pt7}\end{equation}
 Substitute (\ref{pt7}) into (\ref{pt6}) to have \begin{equation}
r_{n}\|f(\hat{\beta})-f(\beta_{0})\|_{2}\le r_{n}\left[\||f_{d}(\beta_{0})\||_{2}\right]\left[\|\hat{\beta}-\beta_{0}\ \|_{2}\right]+o_{p}(1).\label{pt8}\end{equation}
 In (\ref{pt8}) on the right side note that by Condition C2, we have
$r_{n}\|\hat{\beta}-\beta_{0}\|_{2}=O_{p}(1)$. \begin{equation}
r_{n}\|f(\hat{\beta})-f(\beta_{0})\|_{2}\le\left[\||f_{d}(\beta_{0})\||_{2}\right]O_{p}(1)+o_{p}(1).\label{pt9}\end{equation}
 Next, divide each side by $r_{n}$ to have \begin{equation}
\|f(\hat{\beta})-f(\beta_{0})\|_{2}\le\left[\||f_{d}(\beta_{0})\||_{2}\right]O_{p}(\frac{1}{r_{n}})+o_{p}(\frac{1}{r_{n}}).\label{pt10}\end{equation}
 Then parts a)-c) follow through by substituting the different specifications
for $\left[\||f_{d}(\beta_{0})\||_{2}\right]$ on the right side.

Note that we can prove Theorem \ref{thm1} in $l_{1}$ and $l_{\infty}$
norms, too. To see this, (\ref{pt2})-(\ref{pt4}) also hold with
$l_{1}$ and $l_{\infty}$ norms. This is true since Lemma 2.12 of
van der Vaart (2000) holds also with $l_{1}$ and $l_{\infty}$ norms.
Hence, the proof will follow with $l_{1}$ and $l_{\infty}$ versions
of C1 and C2 with (\ref{r2}) and (\ref{r4}).

\textbf{Q.E.D.}\\
 \textbf{Remark}. Note that the proof of Theorem \ref{thm1} mainly
uses triangle and Cauchy-Schwarz inequalities. Under some special
conditions, these inequalities hold with the equality sign. For example,
$|a'b|=\|a\|_{2}\|b\|_{2}$ and $\|a+b\|_{2}=\|a\|_{2}+\|b\|_{2}$
when $a$ and $b$ are on the same ray. However, these conditions
are very restrictive and they do not hold in general. Thus, we only
have inequalities in most cases.

PART C.\\

\textbf{Example AC.1}. Here we use the setup on p.690 and p.698 of
Newey and Windmeijer (2009), which is also used in section 2 of Caner
(2014). For $i=1,\cdots,n$ \[
y_{i}=x_{i}'\beta_{0}+\epsilon_{i},\]
 \[
x_{i}=\Psi_{i}+\eta_{i},\]
 and we have $E[\epsilon_{i}|Z_{i},\Psi_{i}]=0$, $E[\eta_{i}|Z_{i},\Psi_{i}]=0$.
$\eta_{i},\epsilon_{i}$ are correlated. Also $y_{i}$ is a scalar,
$x_{i}$ is a $p\times1$ vector of control and endogenous variables,
and $p$ is fixed here. $Z_{i}$ is an $m\times1$ vector of instrumental
variables. $\Psi_{i}$ is a $p\times1$ vector of reduced form values.
The moment function is an $m\times1$ vector, where $m$ grows with
$n$, for $\beta$ in a compact subset of $R^{p}$: \[
\hat{g}(\beta)=\frac{1}{n}\sum_{i=1}^{n}Z_{i}(y_{i}-x_{i}'\beta).\]
 First, define the sequence $\mu_{n}$ such that $\mu_{n}=o(n^{1/2})$.
This will indicate less than strong identification. Suppose it has
the reduced form \[
x_{i}=(z_{1i}',x_{2i}')',\]
 \[
x_{2i}=\pi_{21}z_{1i}+\frac{\mu_{n}}{n^{1/2}}z_{2i}+\eta_{2i},\]
 \[
Z_{i}=(z_{1i}',Z_{2i}')',\]
 where $z_{1i}$ is a $p_{1}$ vector of included exogenous variables
(controls), $z_{2i}$ is a $(p-p_{1})\times1$ vector of excluded
exogenous variables, and $Z_{2i}$ is an $(m-p_{1})\times1$ vector
of instruments. If $z_{2i}$ is not observed, this is approximated
by $Z_{2i}$ which can be a power series or splines. Define a $p\times p$
matrix \[
\tilde{S}_{n}=\left[\begin{array}{cc}
I_{p_{1}} & 0\\
\pi_{21} & I_{p-p_{1}}\end{array}\right].\]
 Define $S_{n}=\tilde{S}_{n}diag(\mu_{1n},\cdots,\mu_{p_{1},n},\mu_{p_{1}+1,n},\cdots,\mu_{p,n})$
where the $diag()$ is a $p\times p$ diagonal matrix, its first $p_{1}$
elements are $n^{1/2}$, and the rest is $\mu_{n}$. So $\mu_{j,n}=n^{1/2}$
for $j=1,\cdots,p_{1}$ and $\mu_{j,n}=\mu_{n}$ for $j=p+1,\cdots,p$.
The reduced form can be written as: \[
\Psi_{i}=\left(\begin{array}{c}
Z_{1i}\\
\pi_{21}z_{1i}+\frac{\mu_{n}}{n^{1/2}}z_{2i}\end{array}\right)=\frac{S_{n}z_{i}}{n^{1/2}}.\]
 Note that \[
\hat{g}(\hat{\beta})-\hat{g}(\beta_{0})=\frac{1}{n}Z_{i}x_{i}'(\hat{\beta}-\beta_{0}),\]
 as in supplement (Appendix) p.11 of the proof of Theorem 2 in Newey
and Windmeijer (2009). Condition C1 is not needed since the system
is linear in $\beta$. There is no need for (\ref{pt2}). We benefit
from a direct proof of Theorem \ref{thm1}. By Theorem 3 of
Newey and Windmeijer (2009) \begin{equation}
\|S_{n}'(\hat{\beta}-\beta_{0})\|_{2}=O_{p}(1),\label{exa1}\end{equation}
 which is Condition C2. Next with $f_{d}(\beta_{0})=\hat{G}=\frac{-1}{n}\sum_{i=1}^{n}Z_{i}x_{i}'$.
So clearly \[
\|\hat{g}(\hat{\beta})-\hat{g}(\beta_{0})\|_{2}=\|\hat{G}(\hat{\beta}-\beta_{0})\|_{2}.\]

Then
\begin{eqnarray*}
n^{1/2}\|(\hat{g}(\hat{\beta})-\hat{g}(\beta_{0})\|_{2} & = & \|\hat{G}n^{1/2}S_{n}^{-1'}S_{n}'(\hat{\beta}-\beta_{0})\|_{2}\\
 & \le & \|\hat{G}n^{1/2}S_{n}^{-1'}\|_{2}\|S_{n}'(\hat{\beta}-\beta_{0})\|_{2}\\
 & = & O_{p}(1)O_{p}(1),\end{eqnarray*}
 by the proof of Lemma A.7 in Newey and Windmeijer (2009), we have
$\|\hat{G}n^{1/2}S_{n}^{-1'}\|_{2}=O_{p}(1)$, and the other rate
is by (\ref{exa1}). Since $S_{n}$ has mixed rates, $n^{1/2}$ and
$\mu_{n}$, we find that $n^{1/2}$ is faster or equal to them, satisfying
Theorem \ref{thm1}.

\textbf{Example AC.2.}

Example 3 in the main text considers the out-of-sample portfolio variance
using the global minimum portfolio weights. Alternatively, we can
apply Theorem \ref{thm1} to study the convergence rate of the estimated
portfolio variance using an estimated matrix $\hat{\Sigma}$ and a
given weight $w$. Hence, in this example we are interested in the
parameter $\beta_{0}=vech(\Sigma_{0})$ and the function \[
f(\beta_{0})=w^{\prime}\Sigma_{0}w=(w^{\prime}\otimes w^{\prime})D_{p}vech(\Sigma_{0})=(w^{\prime}\otimes w^{\prime})D_{p}\beta_{0},\]
 where $\Sigma_{0}$ is $q\times q$, $w$ is $q\times1$, $\beta_{0}$
is $p_{n}\times1$ with $p_{n}=q(q+1)/2$, and $D_{p}$ is a duplication
matrix. Let $q$ (and thus $p_{n}$) be growing with $n$. Since $f(\beta)$
is linear in $\beta$, Condition C1 is always satisfied. Recall that
$\|A\|_{\infty}=max_{i}\max_{j}|A_{ij}|$ of the matrix A, (note that
this is different than $\||A\||_{\infty}$ which is the maximum row-sum
matrix norm), $\||A\||_{2}$ is the Frobenius norm of matrix A, $\|v\|_{2}$
is the $l_{2}$ norm of a vector $v$.

For Condition C2, van de Geer et al. (2014) we have \[
\|\hat{\Sigma}-\Sigma\|_{\infty}=O_{p}\left(\sqrt{\frac{\log q}{n}}\right)\]
 by the symmetry of $\hat{\Sigma}-\Sigma$, so by p.365 of Horn and
Johnson (2013) \[
\||\hat{\Sigma}-\Sigma\||_{2}\le q\|\hat{\Sigma}-\Sigma\|_{\infty}=O_{p}\left(q\sqrt{\frac{\log q}{n}}\right),\]
 which implies\textbf{ \begin{equation}
\|\hat{\beta}-\beta_{0}\|_{2}\le\||\hat{\Sigma}-\Sigma\||_{2}\le O_{p}\left(q\sqrt{\frac{\log q}{n}}\right).\label{rocex5}\end{equation}
 } \noindent Hence, we require that $q$ has to diverge at a slower rate than
$n$ for C2 to hold.

\noindent Next
\begin{eqnarray*}
\||f_{d}\||_{2} & = & \||(w^{\prime}\otimes w^{\prime})D_{p}\||_{2}=trace[(w^{\prime}\otimes w^{\prime})D_{p}D_{p}^{\prime}(w\otimes w)]^{1/2}\\
 & \le & Eigmax(D_{p}D_{p}^{\prime})^{1/2}\|w^{\prime}\otimes w^{\prime}\|_{2}\\
 & = & \sqrt{2}\|w^{\prime}\otimes w^{\prime}\|_{2}=\sqrt{2}\|w\|_{2}^{2}\le\sqrt{2}\|w\|_{1}^{2},\end{eqnarray*}
 where the third line uses the fact that $Eigmax(D_{p}D_{p}^{\prime})=Eigmax(D_{p}^{\prime}D_{p})=2$
because $D_{p}^{\prime}D_{p}$ is a diagonal matrix with elements
1 and 2. If we assume that $\|w\|_{1}^{2}=k_{n}$, i.e., the gross
exposure is of order $k_{n}^{1/2}$, then \[
f(\hat{\beta})-f(\beta_{0})=w^{\prime}\hat{\Sigma}w-w^{\prime}\Sigma_{0}w=O_{p}\left(k_{n}q\sqrt{\frac{\log q}{n}}\right)\]
 by Theorem \ref{thm1}(b).

Note that by using tools in modern high dimensional portfolio analysis
as in Example 3 with a longer proof, we can get a rate better than
the one in (\ref{rocex5}). Here, we  compare the estimator's rate with the function of the estimator, which is the
portfolio variance estimation by fixed weights. In this example, it
turns out that the gross exposure is the key to the rate. 

\vspace{1em}

Now we provide some perspective in the following remark why our proof technique is desirable.

\textbf{Remark}. This remark shows a  directly derivation of  the bound
for Example 3 in the main text without using technique we propose.
Compared to the proposed upper bound technique, the following alternative method involves more
steps and more comparisons between terms and matrix inequalities.
First, arranging the terms in the out-of-sample variance yields \begin{eqnarray*}
|f(\hat{\beta})-f(\beta_{0})| & = & |\hat{\beta}'\Sigma\hat{\beta}-\beta_{0}'\Sigma\beta_{0}|=|\hat{w}'\Sigma\hat{w}-w'\Sigma w|\\
 & = & |\hat{w}'\Sigma\hat{w}-w'\Sigma\hat{w}+w'\Sigma\hat{w}-w'\Sigma w|\\
 & = & |(\hat{w}-w)'\Sigma\hat{w}+w'\Sigma(\hat{w}-w)|\\
 & = & |(\hat{w}-w)'\Sigma\hat{w}-(\hat{w}-w)'\Sigma w+(\hat{w}-w)'\Sigma w+w'\Sigma(\hat{w}-w)|\\
 & = & |(\hat{w}-w)'\Sigma(\hat{w}-w)+2(\hat{w}-w)'\Sigma w|,\end{eqnarray*}
 where we add and subtract $w'\Sigma\hat{w}$ and $(\hat{w}-w)'\Sigma w$
to get the third and the last equalities, respectively, and we use
symmetricity of $\Sigma$ in the last equality.

 Using the above
equality, we have \begin{eqnarray}
|f(\hat{\beta})-f(\beta_{0})| & = & |(\hat{w}-w)'\Sigma(\hat{w}-w)+2(\hat{w}-w)'\Sigma w|\nonumber \\
 & \le & |(\hat{w}-w)'\Sigma(\hat{w}-w)|+2|(\hat{w}-w)'\Sigma w|\nonumber \\
 & \le & \|\hat{w}-w\|_{1}\|\Sigma(\hat{w}-w)\|_{\infty}+2\|\hat{w}-w\|_{1}\|\Sigma w\|_{\infty}\\
 & \le & \|\hat{w}-w\|_{1}\|\|\Sigma\|_{\infty}\|\hat{w}-w\|_{1}+2\|\hat{w}-w\|_{1}\|\Sigma\|_{\infty}\|w\|_{1}\\
 & = & \|\hat{w}-w\|_{1}^{2}\|\Sigma\|_{\infty}+2\|\hat{w}-w\|_{1}\|\Sigma\|_{\infty}\|w\|_{1},\label{ex3r1}\end{eqnarray}
 where the second line uses the triangle inequality, the third line
uses Holder's inequality, the fourth line follows from $\|Ax\|_{\infty}\le\|A\|_{\infty}\|x\|_{1}$
for a generic $A$ matrix and a generic vector $x$ by our definition
of $\|A\|_{\infty}$.

Next, we evaluate the first term in (\ref{ex3r1}). By Theorem 3.3
of Callot et al. (2019), we have \[
\|\hat{w}-w\|_{1}=O_{p}(\frac{\sqrt{logp_{n}}}{\sqrt{n}}\bar{s}^{3/2}).\]
 We assume then $\|\Sigma\|_{\infty}=O(1)$. So the first term's rate
is: \begin{equation}
\|\hat{w}-w\|_{1}^{2}\|\Sigma\|_{\infty}=O_{p}(\frac{\sqrt{logp_{n}}}{\sqrt{n}}\bar{s}^{3/2}).\label{ex3r2}\end{equation}
 Then since $\|w\|_{1}=O(\bar{s}^{1/2})$ by Remark in Theorem 3.3
of Callot et al.  (2019), the second term on the right side of (\ref{ex3r1})
has the rate \begin{equation}
2\|\hat{w}-w\|_{1}\|\Sigma\|_{\infty}\|w\|_{1}=O_{p}(\frac{\sqrt{logp_{n}}}{\sqrt{n}}\bar{s}^{3/2})O_{p}(\bar{s}^{1/2})=O_{p}(\frac{\sqrt{logp_{n}}}{\sqrt{n}}\bar{s}^{2}).\label{ex3r3}\end{equation}
 So (\ref{ex3r3}) rate is slower than (\ref{ex3r2}), hence it will
be the rate of convergence for the out-of-sample variance estimator.
This means that $r_{n}=\frac{\sqrt{logp_{n}}}{\sqrt{n}}\frac{1}{\bar{s}^{3/2}}$
and $k_{n}=\bar{s}^{1/2}$. By Theorem \ref{thm1}(b), $r_{n}/k_{n}=\frac{\sqrt{logp_{n}}}{\sqrt{n}}\frac{1}{\bar{s}^{2}}$.
Thus, we reach the same conclusion as in Example 3,  but with additional  inequalities and comparisons. Of course,
there is one caveat, the assumptions are slightly different, in example
3 we assume that $\||\Sigma\||_{1}$ is finite, whereas in the remark
here we assume that $\|\Sigma\|_{\infty}$ is finite. In higher dimensions,
the former can be stronger than the latter depending on the sparsity
of columns of $\Sigma$.

Next we provide a remark that measures the divergence between two norms that are used.

\textbf{Remark}. The convergence rate implied by Theorem \ref{thm1}
might be potentially conservative under some cases. Whether the rate
is sharp or conservative depends on the specific setup of the estimator
of interest. To see whether our theorem gives a conservative rate
in Example 3, we can develop a measure of divergence between our upper
bound and the one derived in (\ref{ex3r3}) from the direct proof
in the previous remark. Let $div$ represent the divergence between two
norms of $\Sigma$ \[
div=\frac{\||\Sigma\||_{1}}{\|\Sigma\|_{\infty}}=\frac{\max_{1\le j\le p_{n}}\sum_{i=1}^{p_{n}}|\sigma_{i,j}|}{\max_{1\le j\le p_{n}}\max_{1\le i\le p_{n}}|\sigma_{i,j}|}.\]
 We can see from the divergence measure that our upper bound can be
quite large in certain cases, such as in the case where $\sigma_{i,j}$'s
are constants and bounded away from zero and infinity. But as said
earlier, in case of $\||\Sigma\||_{1}\le C<\infty$, with C being
a positive constant, the rate  is given by our theorem is not conservative.
\end{document}